\documentclass[manuscript]{acmart}

\AtBeginDocument{%
  \providecommand\BibTeX{{%
    \normalfont B\kern-0.5em{\scshape i\kern-0.25em b}\kern-0.8em\TeX}}}

\newcommand{\Tau}{\mathrm{T}}
\newcommand{\shorte}{\textup{\texttt{=}}}

\usepackage{amsthm}
\usepackage{amsmath}

\usepackage{thmtools}
\usepackage{thm-restate}
\usepackage{hyperref}

\allowdisplaybreaks

\usepackage{color}
\usepackage{mathtools}

\usepackage{subfigure}
\usepackage{multirow}
\usepackage{makecell}
\usepackage{array, booktabs, arydshln, xcolor}
\usepackage{graphbox}
\usepackage{tikz}
\usepackage{pgfplots}
\usepackage{wrapfig}

\usepackage{amsmath,amsfonts,bm}

\def\eqref#1{equation~\ref{#1}}

\def\1{\bm{1}}

\def\va{{\bm{a}}}

\DeclareMathAlphabet{\mathsfit}{\encodingdefault}{\sfdefault}{m}{sl}
\SetMathAlphabet{\mathsfit}{bold}{\encodingdefault}{\sfdefault}{bx}{n}

\newcommand{\KL}{D_{\mathrm{KL}}}

\begin{document}

\title{Reinforced Natural Language Interfaces via Entropy Decomposition}


\author{Xiaoran Wu}
\affiliation{%
  \institution{The Department of Computer Science and Technology, Tsinghua University}
  \city{Beijing}
  \country{China}}
\email{wuxr17@tsinghua.org.cn}

\author{Yipeng Kang}
\affiliation{%
  \institution{IIIS, Tsinghua University}
  \city{Beijing}
  \country{China}}
\email{fringsoo@gmail.com}







\renewcommand{\shortauthors}{Xiaoran Wu}

\begin{abstract}
In this paper, we study the technical problem of developing conversational agents that can quickly adapt to unseen tasks, learn task-specific communication tactics, and help listeners finish complex, temporally extended tasks. We find that the uncertainty of language learning can be decomposed to an entropy term and a mutual information term, corresponding to the structural and functional aspect of language, respectively. Combined with reinforcement learning, our method automatically requests human samples for training when adapting to new tasks and learns communication protocols that are succinct and helpful for task completion. Human and simulation test results on a referential game and a 3D navigation game prove the effectiveness of the proposed method.
\end{abstract}



\maketitle

\section{Introduction}

Spoken dialogue interfaces hold the promise to become the main gateways to many human-computer interaction applications. As part of this trend, conversational agents are becoming increasingly popular and integrated into more and more aspects of our lives, providing services like education, retail, and entertainment. However, compared to human-human interactions, existing human-computer dialogue is limited~\cite{luger2016like}, missing the dynamic and interactive nature of the conversation~\cite{serban2016generative, gao2019neural, clark2019makes, seering2019beyond,lazaridou2020emergent}. This discrepancy between the expectation and experience of conversational agents has led to user frustration and the final abandonment of conversational agents~\cite{zamora2017m, amershi2019guidelines}.


These drawbacks are partly attributable to the current training scheme for spoken dialogue interfaces. Typically, agents are passively exposed to corpora consisting of conversation records.
Such a learning paradigm enables the understanding of \emph{structural} properties of natural language, modeling the statistical association between questions and answers/actions of conversational agents. However, conversations become less natural and coherent when encountering dialogues not covered in corpora, limiting the adaptability to new tasks~\cite{ruan2019quizbot, xu2017new}. Moreover, agent policies for action execution tend to be simple and unreliable~\cite{luger2016like}, preventing its application to critical but complex  scenarios, such as human-robot collaboration.

To make up for these drawbacks, conversational agents are expected to communicate real-world information, convey the intentions of humans and agents, and coordinate their cooperative behaviors. These expectations highlight the \textbf{functional} aspect~\cite{wittgenstein1958160} of language as a special coordination system~\cite{zipf1935psycho, grice1975logic, austin1975things, clark1996using} and have revived interests in multi-agent communication. By means  of reinforcement learning, communities of agents must communicate in order to finish a task. Despite much exciting work in this area~\cite{foerster2016learning, sukhbaatar2016learning, das2019tarmac, wang2020learning}, the stress has been on maximizing environmental rewards and finishing the task. Emergent communication protocols are not in natural language and even violate some of its core features, such as Zipf's Law of Abbreviation~\cite{chaabouni2019anti} and compositionality patterns~\cite{kottur2017natural}. Since humans generally do not understand such messages, there are doubts whether this type of functional learning can be used to build conversational human-computer interaction interfaces.

In order to improve the performance of conversational agents, this paper studies the technical problem of building agents that can (1) quickly adapt to new tasks and effectively communicate in natural language with task-specific tactics such as being succinct by dropping words other than "keywords" and (2) help the team (perhaps with humans) finish complex tasks with reliability. Although these expectations cover both the structural and functional aspects of language, we find that they can be achieved simultaneously by minimizing the uncertainty of the spoken language. Mathematically, the uncertainty can be decomposed into an entropy term reflecting the structural uncertainty and a mutual information term measuring whether the message can help the decision-making process of other agents.

The first term helps quick adaption to new tasks. We propose that human samples are requested when the entropy value is large, and these samples are used to minimize an upper bound of the entropy term. In this way, we learn language by minimizing structural uncertainty. The mutual information objective is maximized to make communicated messages contain more information that can reduce uncertainty in other agents' action selection and thus helps policy learning. After optimizing this objective, the value of mutual information is used as a standard for cutting off less important words or phrases in a message.


We demonstrate the effectiveness of our methods on two tasks, a referential game and a 3D navigation task. Experiments show that, for a new task, our approach only requires few human samples to learn task-specific communication protocols in natural language. We design experiments to separately show the effect of the entropy and mutual information learning objective. Furthermore, we carry out human tests to show the ability of our method to communicate and collaborate with humans. We hope that our entropy decomposition approach can provide a theoretical framework for the future design of human-AI spoken dialogue interfaces~\cite{wang2021towards}.

\section{Related Work}

Our research lies at the intersection of human-computer interaction, natural language processing, and (multi-agent) reinforcement learning. In this section, we discuss the relationship with prior works in these three fields.

\subsection{Human-Computer Interaction: Conversational Agents}

When talking about \emph{conversational agents} (CA), one might find some terms synonymous and use them interchangeably~\cite{von2010doesn}, such as chatbot, virtual companion, and autonomous agent. In order to distinguish conversational agents, \citet{wilks2010companion} suggests a series of characteristics. From this perspective, conversational agents are featured by their function -- to carry out tasks. They focus simultaneously on the completion of real-time tasks and developing conversations. In contrast, chatbots only have memory or knowledge about the user to mimic conversation~\cite{li2016deep}. Digital companions also have knowledge about the user but are not designed for any central or overriding tasks~\cite{luger2016like}. Although hundreds of CA platforms have been designed, ranging from social media to forums to gaming platforms, chatbots rarely support interaction among multiple parties~\cite{seering2019beyond} or complex human-robot collaboration tasks. Rather, they focus on task-oriented interactions~\cite{ruan2019quizbot, xu2017new}, leaving a discrepancy between user experience and expectation~\cite{wang2021towards}.

Through a user study, \citet{luger2016like} pinpoint several desired properties of conversational agents. For example, users tend to expect the use of a particular economy of language during interactions, including tactics such as dropping words other than 'keywords,' removing colloquial or complex words, reducing the length of sentences, and using more specific terms. Another concern of users is precision of action execution. Precision is a factor closely linked to trust and reliability. For example, users tend not to use a conversational agent to make a phone call for fear that they will call a wrong person. In this work, we try to make spoken dialogue interfaces more precise via reinforcement learning -- agents learn by trial and error and gradually improve their success rate by learning policies that can maximize the expected return. 

Various topics regarding AI designs for conversational agents are studied, for example, personalization~\cite{findlater2004comparison, gajos2006exploring}, predictability~\cite{gajos2008predictability}, transparency, and trust~\cite{kulesza2015principles, rader2018explanations, weld2018intelligible, ribeiro2016should}. However, existing approaches to designing human-AI interactions lack a theoretical framework~\cite{wang2021towards}.In this work, we provide theoretical support, which is based on entropy decomposition, in the hope that it can improve the interpretability and transparency of spoken dialogue interfaces. 


\subsection{Natural Language Processing: Dialog Systems}
Building conversational agents is a hot topic in the field of natural language processing, where the term \emph{task-oriented dialog systems} is frequently used. Compared with open-domain dialog systems that aim to improve conversational engagement of users, task-oriented dialog systems assist users in finishing one or more specific tasks~\cite{huang2020challenges}, including weather query, restaurant booking, and flight booking.

Currently, two paradigms are used to build task-oriented dialog systems, pipeline and end-to-end methods. In the pipeline paradigm, the system is divided into four modules, a natural language understanding module that maps user utterance to structured semantic representation~\cite{chen2019bert, castellucci2019multi}, a dialog state tracking module that estimates the user's goal~\cite{mrkvsic2017neural}, a dialog policy that generates a high-level system action~\cite{peng2018deep}, and a natural language generation module mapping system actions to natural language utterance~\cite{wen2015semantically}. In contrast, end-to-end methods~\cite{liu2017end, wen2017network, lei2018sequicity} directly output a response in natural language conditioned on the input of a natural language context and are more related to our work. 

Conversations between a dialog system and a user simulator are simulated for the training of end-to-end methods. Reinforcement learning~\cite{dhingra2017towards, shah2018bootstrapping, zhao2019rethinking} and powerful transformers such as GPT-2~\cite{peng2020soloist, hosseini2020simple} have been used to optimize the dialog system. These methods typically require a large task-specific corpus to train policy over conversational actions, while our method only needs few interactions with humans for a new task, and our policies can be extended to actions beyond conversational ones.

Various methods have been proposed to build user simulators. Early work adopts rule-based simulation such as the agenda-based user simulator~\cite{schatzmann2009hidden, li2016user}. Rule-based simulation has to be redesigned for different tasks, limiting the transfer to new tasks. Data-driven modeling improves user simulation by learning language models. \citet{asri2016sequence, gur2018user} use a seq2seq model to generate user conversational actions. The neural user simulator~\cite{kreyssig2018deep} directly generates natural language but learns user behavior entirely from a corpus, preventing the generated conversation from being diverse. Reinforcement learning has also been explored for the user simulator~\cite{tseng2021transferable, takanobu2020multi}. Compared to these methods, our agents further use communication to complete real-world tasks, highlighting the functional aspect of communication. Another difference is that our method can learn a task-aware economic of language for multiple tasks. 

\subsection{Reinforcement Learning: Multi-Agent Communication} 
Cooperative multi-agent reinforcement learning (MARL) aims to learn policies for a team of multiple autonomous agents. The learning objective is to maximize the expected return gained by the team. The environment is typically partial observable, and agents need to communicate to eliminate their uncertainty about the true global state. Previous work learns communication by using differentiable communication channels~\cite{sukhbaatar2016learning, hoshen2017vain, jiang2018learning, singh2019learning, das2019tarmac}. \citet{singh2019learning} propose to learn gates to control the agents to only communicate with their teammates in mixed cooperative-competitive environments. \citet{zhang2013coordinating, kim2019learning} study action coordination under a limited communication channel. Social influence~\citep{jaques2019social} and InfoBot~\citep{goyal2019infobot} propose to penalize messages that have no influence on policies of other agents. TarMAC~\citep{das2019tarmac} uses an attention mechanism to distinguish the importance of incoming messages. These methods put their stress on the functional aspects of communication. Essentially, they learn agents that can develop a protocol sufficient to solve the task at hand, but hoping that the learned protocol possesses desirable characteristics of natural language is wishful thinking~\cite{lazaridou2020emergent}. In contrast, research on pragmatics, e.g., scalar implicatures \citep{rothschild2013game}, politeness implicatures \citep{clark2011meaningful}, irony, debating \citep{glazer2006game}, negotiation \citep{cao2018emergent}, referring expression generation \citep{orita2015discourse}, rhetoric phenomena \citep{kao2014formalizing, kao2014nonliteral}, and realistic human pragmatic behavior \citep{smith2013learning, khani2018planning, achlioptas2019shapeglot, tomlin2019emergent, cohn2018pragmatically, cohn2019incremental}, largely ignores the functional aspects of communication.

\section{Problem Formulation}

In this paper, we aim to let multiple agents learn task-aware communication in natural language to solve a set of tasks $\mathcal{T}$. When deployment, each agent can be replaced by a human and they are expected to finish the task seamlessly. Each task $G \in \mathcal{T}$ can be modelled as a Decentralized Partial Observable Markov Decision Process (Dec-POMDP)~\citep{oliehoek2016concise} $\langle I, S, A, P, R, \Omega, O, n, \gamma\rangle$. Here, $I$ is the set of $n$ agents. $s\in S$ is the true state of the environment from which each agent $i$ draws a partial observation $o_i\in \Omega$ according to the observation function $O(s, i)$. Each agent has an action-observation history $\tau_i\in \Tau\equiv(\Omega\times A)^*$. Since agents only have a partial observation of the environment, they communicate to exchange necessary information to cooperate. Agents learn a language encoder $f$ and send messages to other agents. At each timestep, each agent $i$ receives messages $m_i^{in}$, selects an action $a_i\in A$, forming a joint action $\va \in A^n$, resulting in a shared reward $r=R(s,\va)$ and the next state $s'$ according to the transition function $P(s'|s, \va)$. Each agent learns a policy $\pi_i(\cdot|\tau_i, m_i^{in})$ conditioned on local action-observation history and received messages for action selection. The joint policy $\bm{\pi}=\langle\pi_1,\cdots,\pi_n\rangle$ induces a joint action-value function: $Q_{tot}^{\bm{\pi}}(\bm\tau, \va)=\mathbb{E}_{s_{0:\infty},\va_{0:\infty}}[\sum_{t=0}^\infty \gamma^{t}r_t|s_0\shorte s,\va_0\shorte \va,\bm{\pi}]$, where $\bm\tau$ is the joint history and $\gamma\in[0, 1)$ is a discount factor. Agents need to maximize $Q_{tot}^{\bm{\pi}}$ by learning both policies and communication protocols.

This problem formulation can be used to describe various HCI applications, such as flight booking, weather query, and more complex human-robot cooperation tasks, such as cooperative furniture ensemble. In this paper, demonstrate our approach specifically on a referential game and a 3D navigation task.

For example, in a referential game, a speaker agent and a listener agent observe a target image and a distracting image. Local observation of both agents is $o_i=\langle c, c'\rangle$ where $c$ and $c'$ is the target and distracting image, respectively. Only the speaker knows which image is the target. We expect the speaker to describe the difference between these two images in a succinct way, so that the listener, upon receiving the message, can  distinguish the target image. The listener's action space contains two actions, selecting each of the image as the target. If the listener makes the right decision, both agents are rewarded $0.1$, otherwise $0$. Detailed task settings of the navigation task are provided in Sec.~\ref{sec:exp_house}.
\begin{figure}
    \centering
    \includegraphics[width=\linewidth]{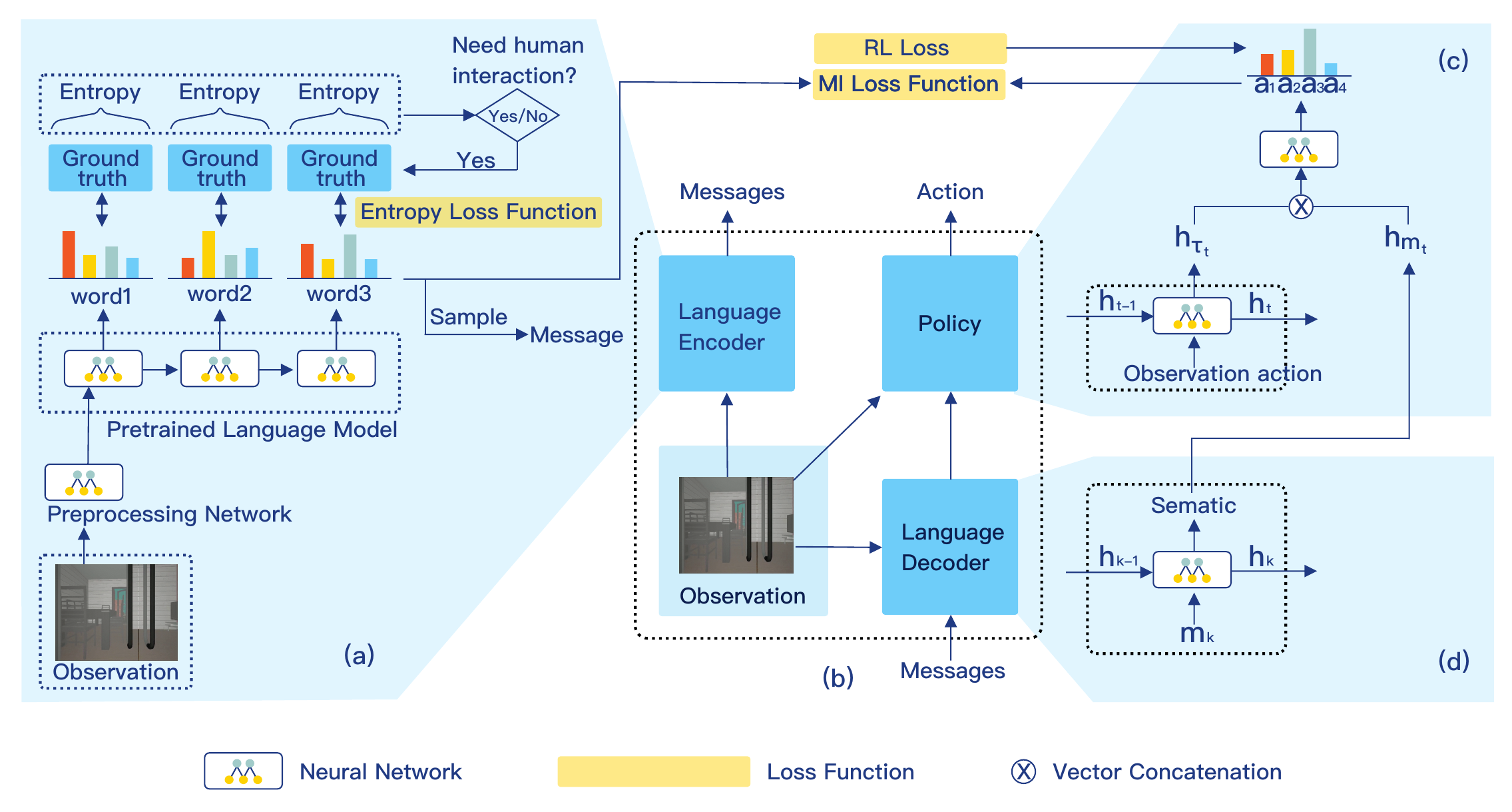}
    \caption{Learning framework of our method.}
    \label{fig:framework}
\end{figure}

\section{Method}
In this section, we introduce our learning framework (Fig.~\ref{fig:framework}). As shown in Fig.~\ref{fig:framework} (b), each agent has three modules, a language encoder for generating messages, a language decoder to process received messages, and a policy for generating action to execute in the environment. 


The language encoder is responsible for generating language to communicate and all our expectations of the learned language depend on this part. Therefore, it is the core component of our method, and we first describe the structure and learning scheme of the language encoder. 

\subsection{Learn Language Encoder via Entropy Decomposition} \label{sec:le}
We expect the language encoder to have the following characteristics. (1) When learning a new task, automatically detect whether it can describe task information in natural language. If not, actively require human interaction for learning. (2) For a given task, the encoder can learn a particular economy of language, such as dropping words other than keywords. (3) The learned language should help agents or humans to finish the task. 

These desiderata may seem daunting at first glance because they cover both functional and structural aspects of language and seem unrelated to each other. Perhaps surprisingly, we show that they can be fulfilled simultaneously by decomposing and minimizing an entropy objective that measures the uncertainty of spoken messages. 

Formally, the language encoder learns $p(m_{ij}^k|m_{ij}^{<k},\tau_i)$ (without loss of generality, we consider the message sent from agent $i$ to $j$) which generates a probability distribution of the $k$th word in the message sent from agent $i$ to $j$, given words before it and the local action-observation history (Fig.~\ref{fig:framework} (a)). We model the language learning problem as the following conditional entropy minimization problem:
\begin{equation}
    \text{minimize}\ H_{\theta_e}(M_{ij} |A_j, \Tau_i).\label{equ:obj}
\end{equation}
where $\theta_e$ is parameters of the language encoder, $M_{ij}$, $A_j$, and $\Tau_i$ is the random variable for messages from $i$ to $j$, actions of agent $j$, and local action-observation history of agent $i$, respectively. Intuitively, Eq.~\ref{equ:obj} is a natural formulation for language learning, because local action-observation history can be regarded as the "cause" for communication while the listener's action is the "result" of communication. Given "cause" and "result", there should be little uncertainty about the messages.

Formally, to see how minimizing Eq.~\ref{equ:obj} helps both structural and functional language learning, we expand it as:
\begin{align}
    H(M_{ij} & |A_j, \Tau_i) = -\int\int\int p(m_{ij}, a_j, \tau_i) \log \frac{p(m_{ij}, a_j, \tau_i)}{p(a_j, \tau_i)}  dm_{ij}da_jd\tau_i \\
    & = -\int\int p(m_{ij}, \tau_i)\log\frac{p(m_{ij}, \tau_i)}{p(\tau_i)} dm_{ij}d\tau_i  - \int\int\int p(m_{ij}, a_j, \tau_i) \log \frac{p(m_{ij}, a_j | \tau_i)}{p(m_{ij}| \tau_i)p(a_j| \tau_i)}  dm_{ij}da_jd\tau_i.
\end{align}
In this way, we can see that the entropy objective is decomposed into two terms:
\begin{equation}
    H(M_{ij} |A_j, \Tau_i) = H(M_{ij} |\Tau_i) - MI(M_{ij}; A_j| \Tau_i),\label{equ:decompose_ent}
\end{equation}
where the first term is an entropy conditioned only on local action-observation history and the second term is the mutual information between the message and the listener's action selection. Next, we show how optimizing these two terms can achieve our expectation of the generated language. 

\textbf{Structural learning.}\ \ We expand the entropy term $H(M_{ij} |\Tau_i)$ as:
\begin{align}
    H(M_{ij} |\Tau_i) = \mathbb{E}_{m_{ij},\tau_i} \left[\log p(m_{ij} | \tau_i)\right] = \sum_{k} \mathbb{E}_{m_{ij},\tau_i} \left[ \log p(m^k_{ij} | m^{<k}_{ij}, \tau_i)\right] = \sum_{k} H(M_{ij}^k | M_{ij}^{<k}, \Tau_i),\label{equ:structive_obj}
\end{align}
where $m^k_{ij}$ is the $k$th word in $m_{ij}$ and $m^{<k}_{ij}$ is the words before it. This entropy reflects the uncertainty of a word generated given previous words and local information, and thus serves the role of a language model (estimating $p(m^k|m^{<k}, \tau_i)$~\cite{bengio2003neural}), which measures the structural quality of the generate language. We now discuss how to minimize this term.

The simplest way to optimize Eq.~\ref{equ:structive_obj} is to treat the entropy as a loss and update parameters in the language encoder using stochastic gradient descent. Although this method can reduce the uncertainty of generated messages, it holds no guarantee that the learned messages would be in natural language. To solve this problem, we propose to minimize an upper bound of Eq.~\ref{equ:structive_obj}:
\begin{equation}
    \mathcal{CE}\left[p(M_{ij}^k | M^{<k}_{ij}, \Tau_i) \| q(M_{ij}^k | M^{<k}_{ij}, \Tau_i)\right] = H(M_{ij}^k | M_{ij}^{<k}, \Tau_i) + \KL\left[p(M_{ij}^k | M^{<k}_{ij}, \Tau_i) \| q(M_{ij}^k | M^{<k}_{ij}, \Tau_i)\right].
\end{equation}
Here, $\mathcal{CE}[\cdot\|\cdot]$ is the cross entropy operation, and $\KL[\cdot\|\cdot]$ is the KL divergence. Since the KL divergence is always larger than 0 for any distribution $q$, we can minimize the cross entropy term as an upper bound in order to minimize the entropy $H(M_{ij}^k | M_{ij}^{<k}, \Tau_i), \forall k$.

The advantage of introducing this upper bound is that we can request human participation, asking them what they will say when observing a certain $\tau_i$, which acts as ground truth labels for training our language encoder $p(M_{ij}^k | M^{<k}_{ij}, \Tau_i)$. In practice, when the entropy $H(M_{ij}^k | M_{ij}^{<k}, \Tau_i)$ is larger than a threshold $t_H$ (we discuss the influence of the value of $t_H$ in Sec.~\ref{sec:exp_structural}), we require human participation and carry out supervised training by minimizing the cross entropy.

We expect that we only need a few human interactions for training a new task. To this end, we additionally use a pre-trained language model (Fig.~\ref{fig:framework} (a)) to initialize our language encoder. Different from previous work on conversational agents, this language model can be pre-trained on any corpus which is not necessarily related to the tasks in hand. In this way, for learning communication in a new task, we only require a few human annotations to align the pre-trained language model with the task semantics. 

\textbf{Functional learning.}\ \  We now consider the second term in Eq.~\ref{equ:decompose_ent}. When minimizing the entropy $H(M_{ij} |A_j, \Tau_i)$, we are maximizing this mutual information objective. The mutual information $MI(M_{ij} ; A_j| \Tau_i)$ measures how much uncertainty about the listener's action selection $A_j$ can be reduced after message $M_{ij}$ is given, which exactly reflects the functional aspect of communication -- agents communicate to reduce the uncertainty of each other's decision-making.

To see how we can maximize the mutual information objective, we expand it as:
\begin{align}
    MI(M_{ij} & ; A_j| \Tau_i) = \sum_k \int\int\int p(m_{ij}^{\le k}, a_j, \tau_i) \log \frac{p(m_{ij}^{k} |m_{ij}^{<k}, a_j, \tau_i)}{p(m_{ij}^{k} | m_{ij}^{<k}, \tau_i)} dm_{ij}da_jd\tau_i\\ 
    = & \sum_k \int\int\int p(m_{ij}^{\le k}, a_j, \tau_i) \log \frac{p(a_j | m_{ij}^{\le k})}{p(a_j | m_{ij}^{<k})} dm_{ij}da_jd\tau_i= \sum_k \mathbb{E}_{m_{ij}, \tau_i}\left[ \KL[p(A_j | M_{ij}^{\le k})\| p(A_j| M_{ij}^{<k})]\right]. \label{equ:functional_obj}
\end{align}
Similar to Eq.~\ref{equ:structive_obj}, $m_{ij}^{k}$ is the $k$th word in $m_{ij}$, and $m_{ij}^{<k}$ is the words before it. Intuitively, Eq.~\ref{equ:functional_obj} measures how much the listener's policy distribution differs with and without the knowledge of words in $m_{ij}$.

The KL divergence involves two distributions over the listener's action, thus directly optimizing it with respect to the parameters of the language encoder requires a differential communication channel to enable gradients to flow from the listener to the speaker, which is not a natural assumption. Therefore, we propose to treat the KL divergence as a reward and use the policy gradient theorem~\cite{sutton2018reinforcement} to optimize the language encoder. Eq.~\ref{equ:functional_obj} also decomposes the mutual information term into contribution of each word in the sentence, enabling us to use the following gradients for a given $(m_{ij}^{< k}, m_{ij}^{k}, \tau_i)$ tuple:
\begin{equation}
    g_{MI}(\theta_e) = \sum_k \KL\left[p(A_j | m_{ij}^{\le k})\| p(A_j| m_{ij}^{<k})\right] \nabla_{\theta_e} p(m_{ij}^{k} | m_{ij}^{< k}, \tau_i).\label{equ:functional_gradient}
\end{equation}

\subsection{Language Decoder and Policy}
Upon receiving messages from other agents, the language decoder processes the messages by a recurrent neural network (RNN; we use GRU~\cite{cho2014learning} in practice), parameterized by $\theta_d$. At each RNN step, we input a word in the message and get a vector representing the message's semantic meaning. The policy module, with parameterization $\theta_p$, also uses a RNN which processes local action-observation history. At each timestep $t$, the RNN consumes observation $o_t$ and the action at the last timestep $a_{t-1}$. The output of the RNN module is concatenated with the semantic vector and is fed into a fully-connected network to generate action probability distribution $\pi(a|\tau, m^{in})$. Here, we omit the subscript $i$ because this structure is the same for every agent. The language decoder and the policy module are trained end-to-end via REINFORCE~\cite{williams1992simple} policy gradients:
\begin{equation}
    g(\theta_d, \theta_p) = \nabla_{\theta_d, \theta_p}\mathbb{E}_{(\tau, a)\sim\pi}\left[ R(\tau, a) \log \pi(a|\tau, m^{in})\right]
\end{equation}
Here, $R(\tau, a)$ is the accumulated rewards gained in the episode after taking action $a$ under local action-observation history $\tau$. Intuitively, this optimization rule updates parameters of the language decoder and policy so that the probability of taking more rewarding actions are increased.
\section{Experiments}\label{sec:exp}
In this section, we conduct experiments to answer the following questions. (1) For structural language learning, we minimize the conditional entropy of the messages given local action-observation history. The first question is whether this method can quickly learn suitable language on a new task. (2) According to the value of the conditional entropy (Eq.~\ref{equ:structive_obj}), we request human interactions to provide supervised learning labels. The question is how the amount of human interactions influences the learning outcome. (3) For functional language learning, we maximize the mutual information between messages and the listener's action selection. Whether this optimization can help improve the precision of the learned language? (4) We are curious about whether the value of mutual information can be used to identify "keywords" in a sentence. (5) How does the mutual information objective, conditional entropy objective, and the reinforcement learning objective affect the learning process? (6) Can the learned agents communicate and collaborate smoothly with human experimenters? 

In the following sections, we first introduce our experimental settings and then show the results to answer the above questions. We carry out experiments on NVIDIA 2080Ti GPUs, the training phase takes about 3 hours.

\subsection{Experiment Environments}
We demonstrate that our method can adapt quickly to new tasks and learn language with desired properties in two tasks, a referential game and a 3D navigation game. 
\begin{figure}
    \centering
    \includegraphics[width=0.95\linewidth]{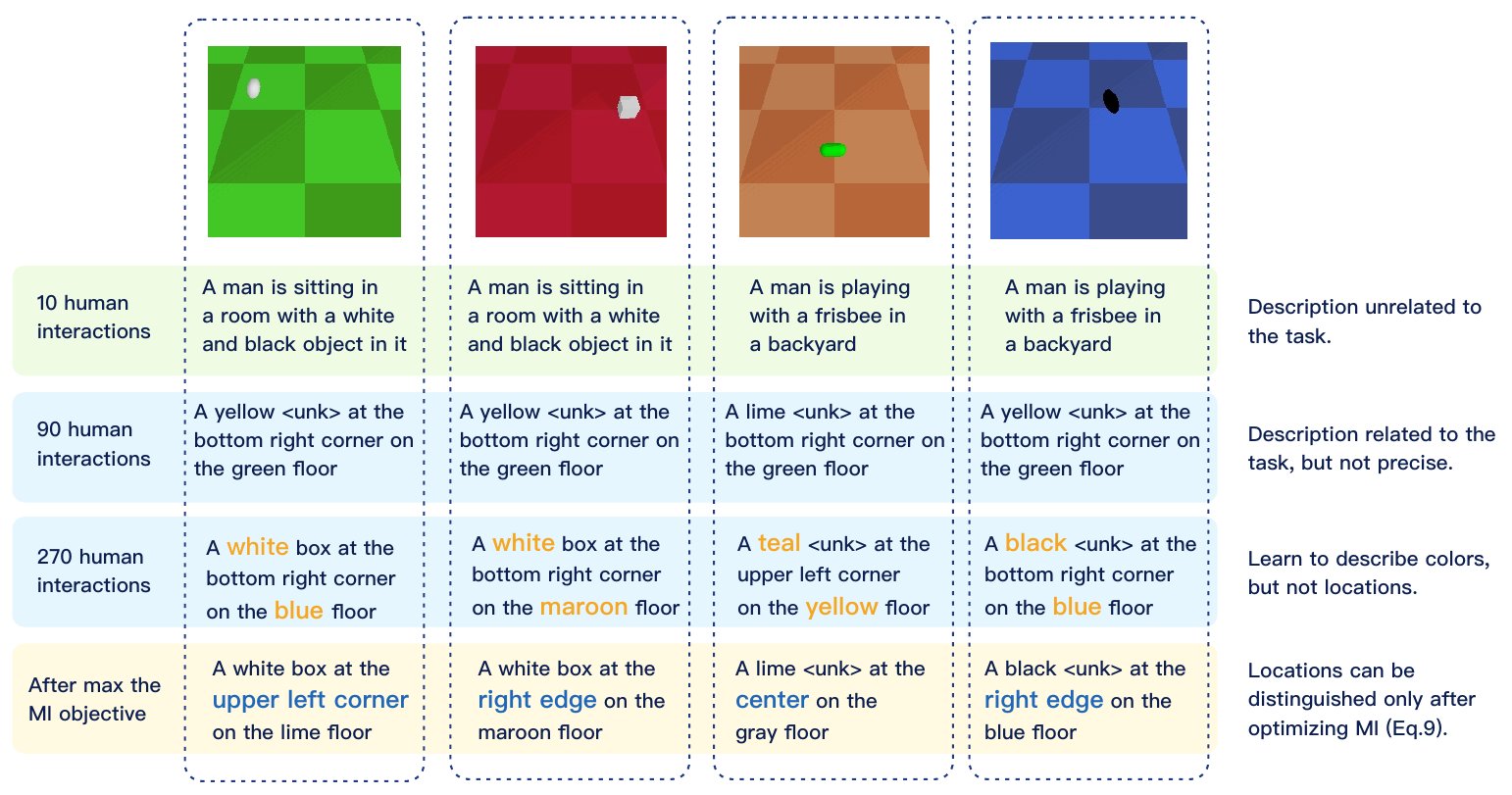}
    \caption{Learned language for the referential game, after supervised training using different numbers of human samples, and after optimizing the mutual information objective (Eq.~\ref{equ:functional_gradient}).}
    \label{fig:captions}
\end{figure}

\subsubsection{Referential game}
There are two agents in this game, a speaker and a listener. Both of them are shown two images, one target and one distracting image. Only the speaker knows which is the target. We expect that the speaker can use succinct language to help the listener distinguish the target from the distracting image. We generate images using PyBullet~\cite{coumans2016}, containing MuJoCo-like objects with different colors, shapes, and locations on a color-changing background. The training set consists of 9,000 such images and the test set contains 1,000. For training, images are repeatedly sampled as target or distractor. We train the agents for 20K rounds of game. Note that agents have different viewpoints on the target, so that they do not learn to communicate simply by comparing pixel-wise information. Instead, they are expected to learn to encode the object features in natural language and then decode them for policy learning. In order to reduce the training time, we initialize our model with a pretrained ResNet to preprocess the images. To encourage reinforcement learning exploration, agent actions are sampled from $\pi$, and these distributions are penalized by an entropy regularization term. We allow messages to have a maximum length of 15 words, and the vocabulary size is 9490. For a testing phase, each of the 1,000 test images serves as the target once, while the distractor in each round is sampled randomly from the test set.

\subsubsection{Navigation Game in 3D Scenarios}\label{sec:exp_house}
In this game, two agents navigate in a house, and we require them to enter a target room simultaneously to get rewarded. If one of them reaches the room earlier, the episode ends and agents will be punished. We create our environment based on the House3D~\cite{wu2018building} simulator, which consists of 45,622 human-designed 3D scenes ranging from single-room studios to multi-floor houses. 

Each scene in House3D is annotated with 3D coordinates and its room and object types (e.g. bedroom, table, etc). The observation space of this task consists of the following signals: (1) the visual signal of an agent's first person view (RGB images); (2) an instance segmentation mask distinguishing all the objects visible in the current view; (3) depth image. These settings ease the preprocessing of sensor inputs and let us focus on the language and policy learning parts. The designers also provide top-down 2D maps, connectivity analysis and shortest paths between two points. 

\subsection{Structural Language Learning}\label{sec:exp_structural}
\begin{figure}
    \centering
    \subfigure{\includegraphics[width=0.32\linewidth]{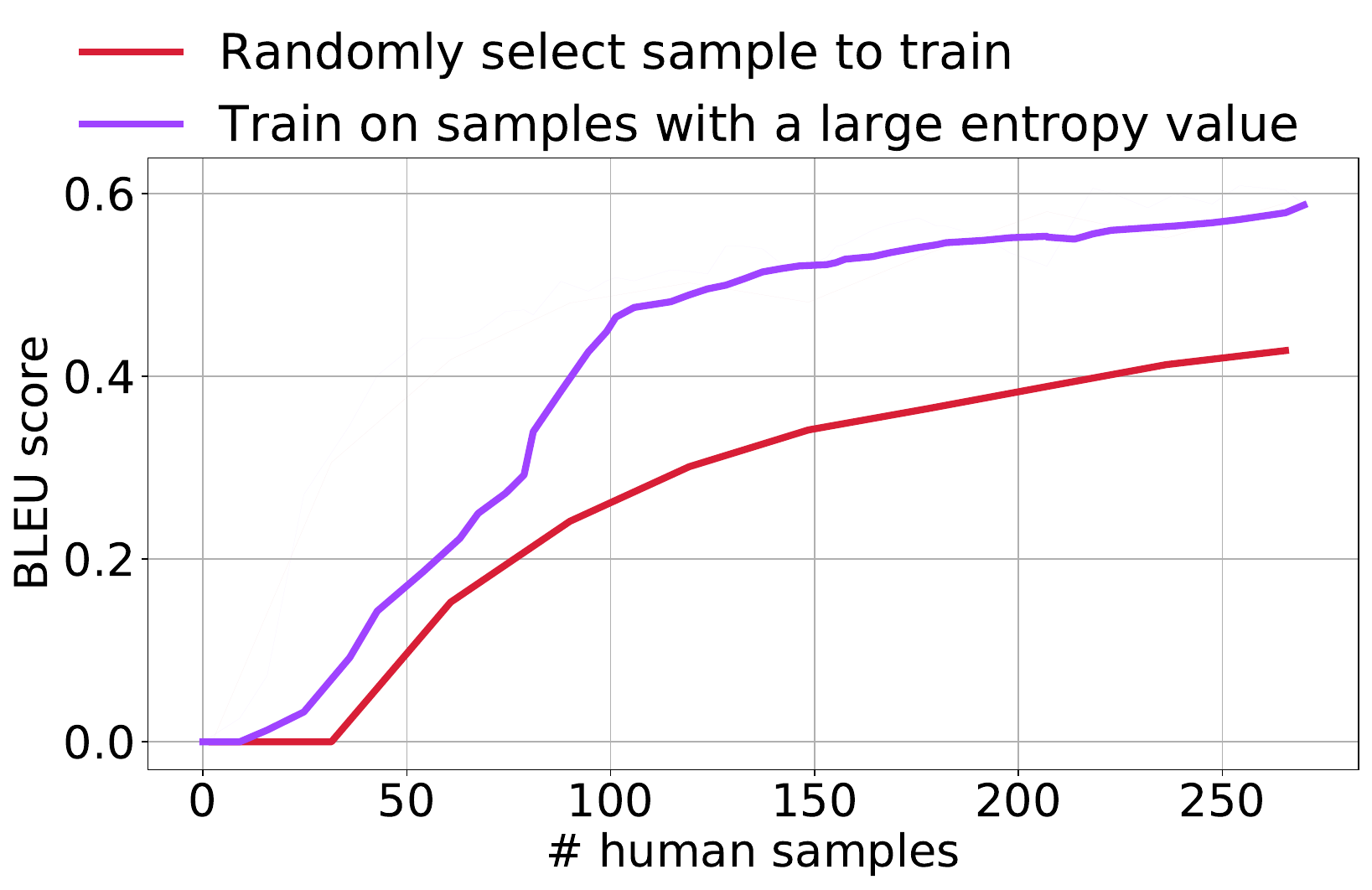}}
    \subfigure{\includegraphics[width=0.32\linewidth]{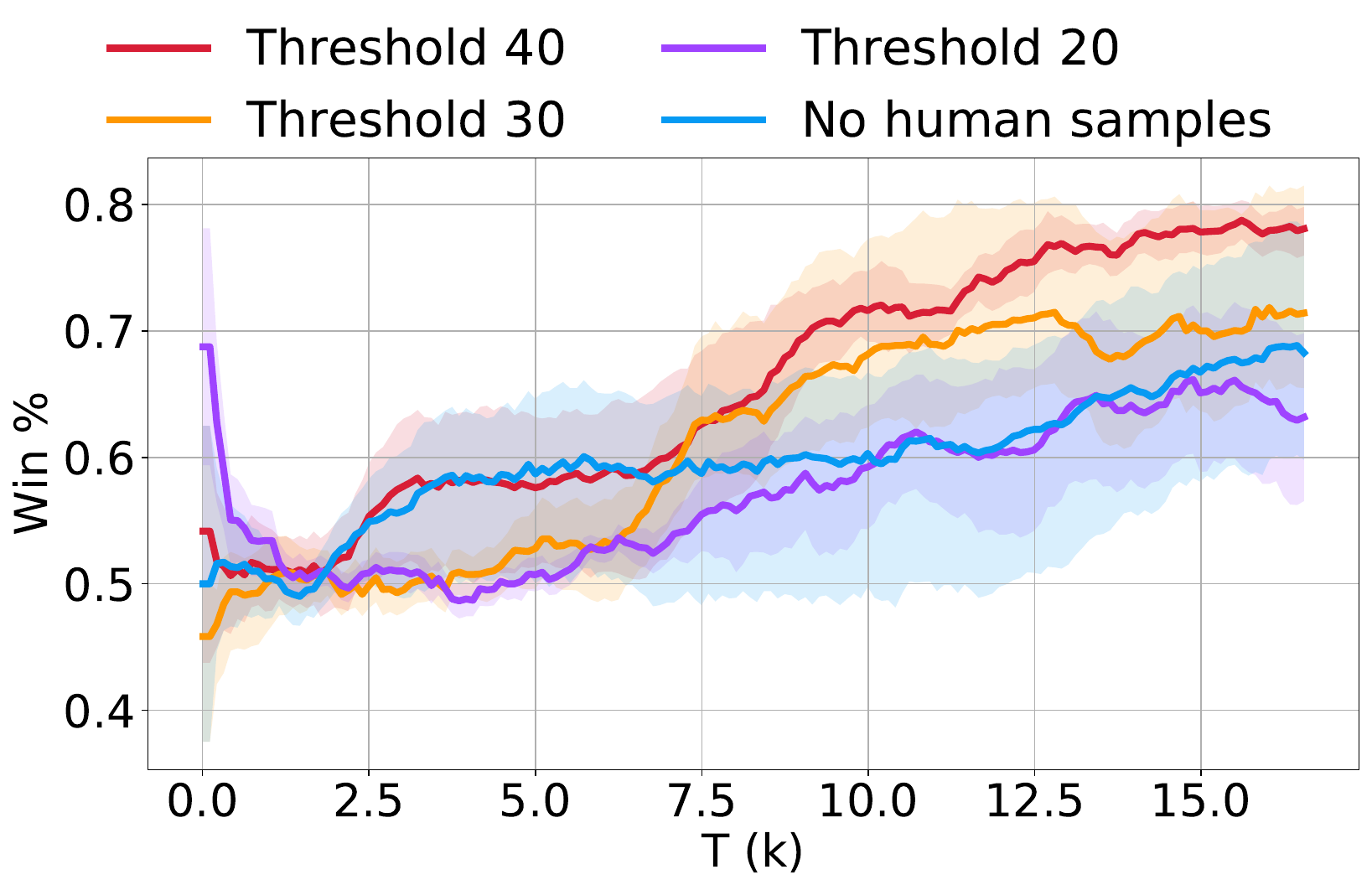}}
    \subfigure{\includegraphics[width=0.32\linewidth]{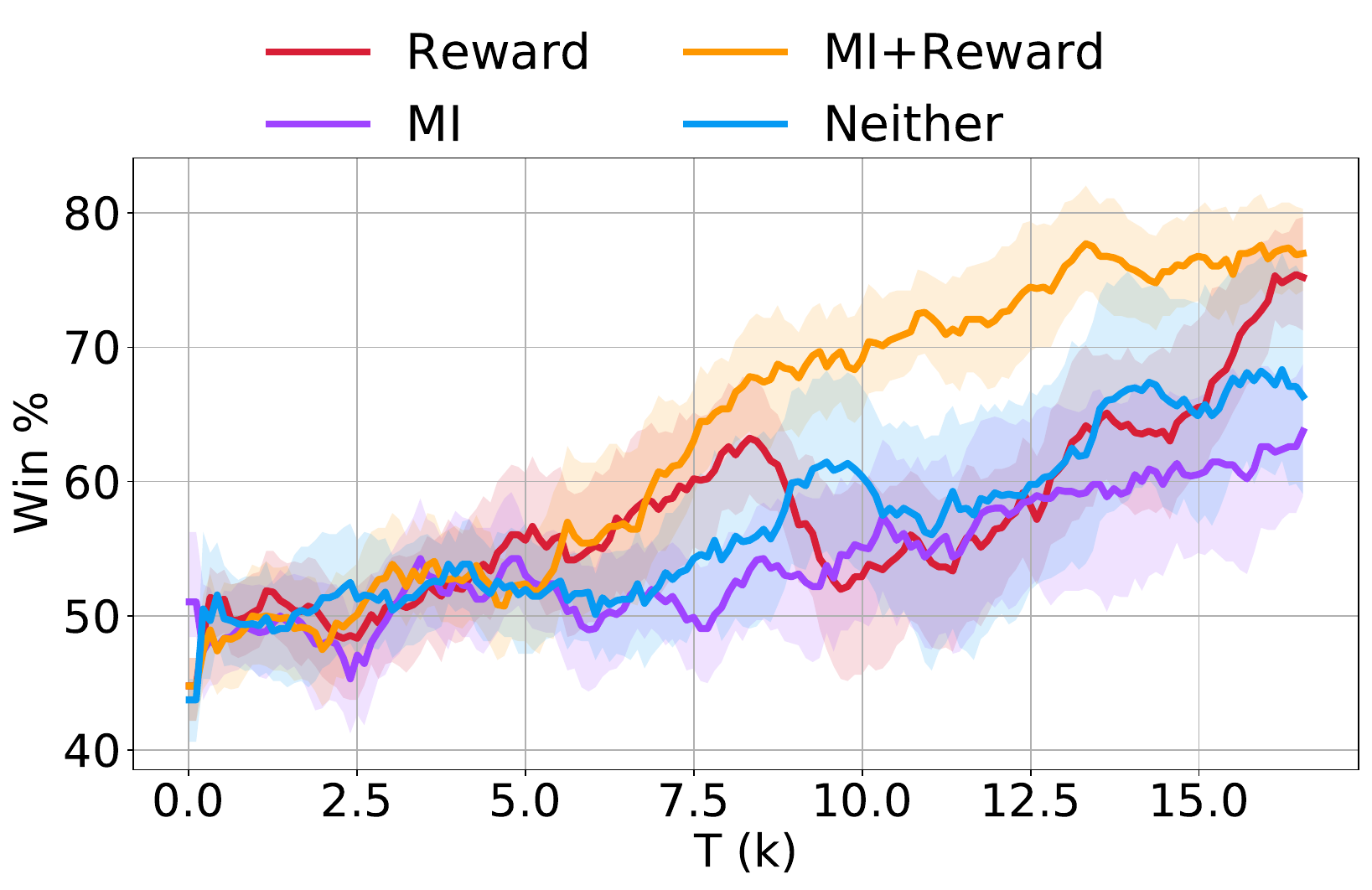}}
    \caption{(a) Requesting human samples according to the value of entropy learns more natural language than random requests. (b) The influence of different entropy thresholds on learning performance. (c) Optimizing the mutual information loss accelerates learning. }
    \label{fig:comp-1}
\end{figure}
\textbf{\emph{Human samples and the learned language.}} We calculate the entropy $H(M_{ij}|O_i)$ and request human annotations when its value is larger than a threshold. Human annotations are then used for supervised learning to train the language encoder. To demonstrate the effect of this learning objective, for the referential game, we show the learned communication protocols in Fig.~\ref{fig:captions}.

\begin{wrapfigure}[12]{r}{0.4\linewidth}
    \centering
    \includegraphics[width=\linewidth]{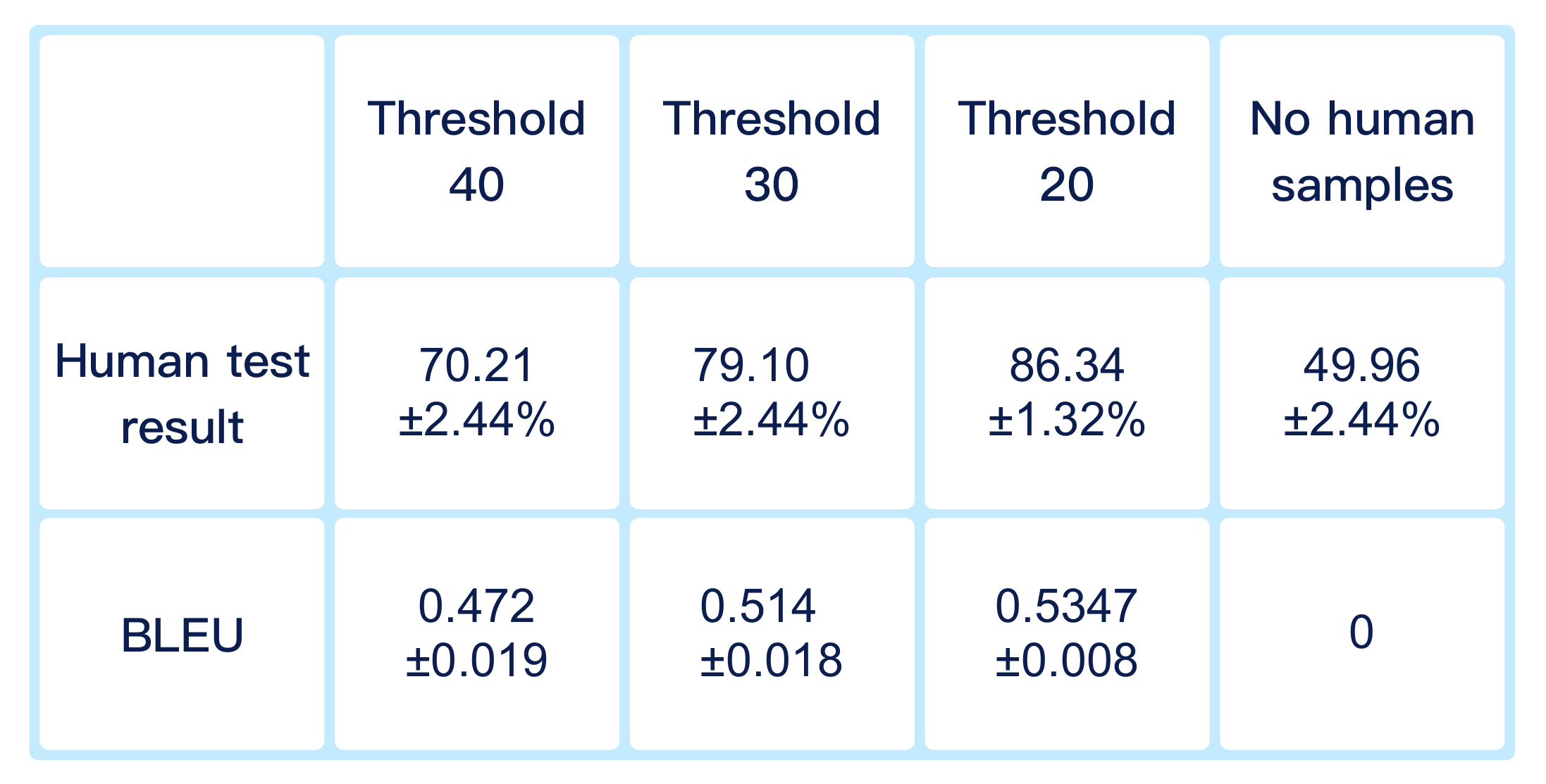}
    \caption{Up: the success rate of the referential game when we replace the listener with human. Bottom: BLEU scores of the learned language encoder. Results under different entropy thresholds are shown.}
    \label{tab:human_test}
\end{wrapfigure}

Since the pre-trained language model is learned on a general corpus that is unrelated to the referential game, initial descriptions seem arbitrary and contain little distinguishing information for the target image. For example, it describes both the first and second image as "A man is sitting in a room with a white and black object in it". One can image that such a language model can hardly collaborate with humans, as proven by our human test results shown in Fig.~\ref{tab:human_test}.

The language becomes related to the task after trained with 90 samples provided by human. However, protocols at this time still performs poorly on the referential game. It cannot distinguish critical features such as color and location of the object in the image. With more human annotations provided, the precision of language improves. After supervised training with 270 human samples, the language model is able to describe the color of the object and floor.

We observe that distinguishing colors is the upper limit of the ability of the supervised learning objective. Providing more samples will not improve the language encoder. We hypothesize that this is because color is such a feature that can be easily captured, but not locations.

\textbf{\emph{Request human samples randomly vs. according to the value of entropy.}} Fig.~\ref{fig:captions} shows that human samples enable the language encoder to use task-related messages which can convey some important information for the task. However, whether it is good to require human samples according to the value of the conditional entropy ($H(M_{ij}^k | M_{ij}^{<k}, \Tau_i)$, Eq.~\ref{equ:structive_obj}) is still unclear. To make up for this drawback, we train the language encoder by randomly selecting samples (with a probability of $20\%$) and require human annotations for them. We compare the BLEU score (a number between 0 and 1 measuring the similarity of the generate language with the ground truth) of the random schedule and the proposed schedule in Fig.~\ref{fig:comp-1} (a). We can see that, with the same number of human samples, selecting samples according to the value of entropy achieves a higher BLEU score, consolidating our theoretical analysis. 

\textbf{\emph{Influence of entropy threshold}} To study the influence of different thresholds for requesting human samples, we test three thresholds (20, 30, 40) and show the learning curves in Fig.~\ref{fig:comp-1} (b). Results in this figure are obtained by carrying out experiments with 3 random seeds. The mean success rate and $95\%$ confidence intervals are shown. We can conclude that the larger the threshold is, the higher win rate can be achieved. For example, When the threshold is set to 40, a mean success rate of 0.77 can be achieved. 

We further present the BLEU scores of the best model obtained under different thresholds. Perhaps interestingly, the larger the threshold is, the lower the BLEU score. This observation indicates that more human samples can make the learned protocols possess more structural features of natural language, but will hurt the reinforcement learning performance. We hypothesize that this is because when the threshold is low, the language encoder is trained for a longer time. Therefore, from the perspective of the listener, even when the observations are the same, the received messages can be different. This introduces extra non-stationarity~\cite{tan1993multi} into reinforcement learning process, rendering slow and unstable convergence.

\textbf{\emph{Human experiments}} To demonstrate that our method can enable fluent human-computer interaction, we replace the reinforcement learning listener with human experimenters. We can see that humans also cannot distinguish the target image from the distracting image when no human samples are used. Moreover, when more human interactions are collected, human experimenters tend to perform better, which is in line with the indication of higher BLEU scores. 

\begin{figure}
    \centering
    \includegraphics[width=\linewidth]{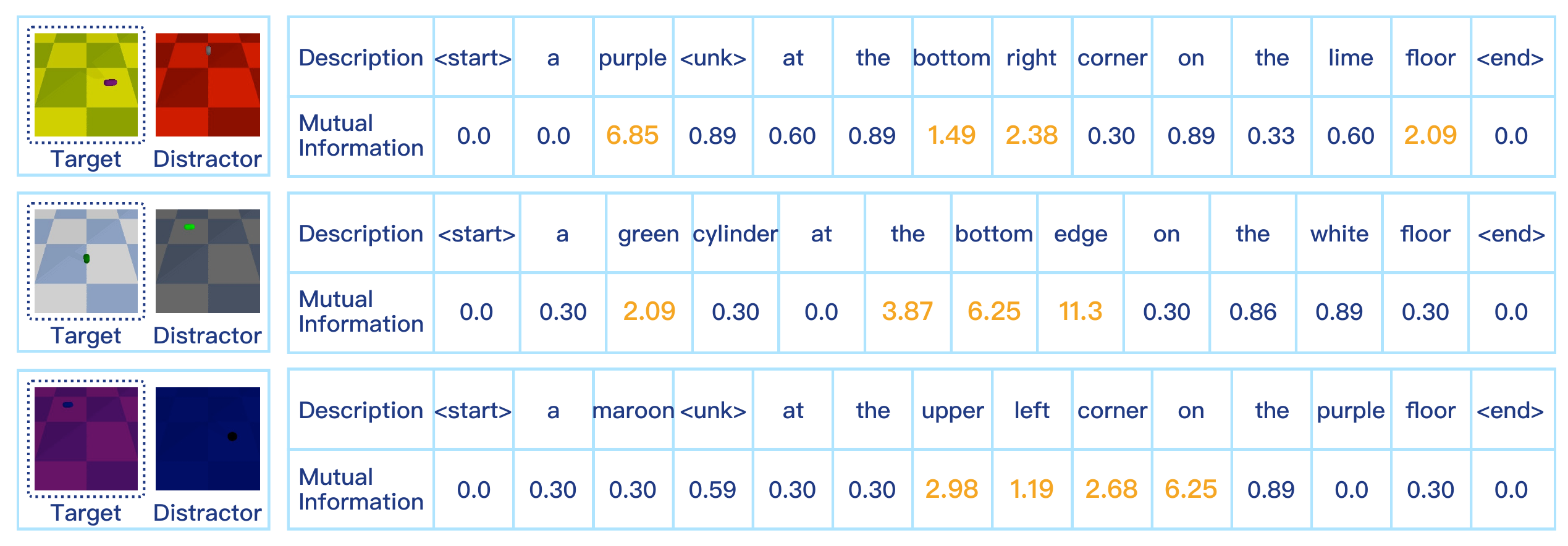}
    \caption{Mutual information between each word and the listener's action selection in the referential game. }
    \label{fig:word_mi}
\end{figure}
\subsection{Functional Language Learning}
We propose to maximize the mutual information between messages and the listener's action selection to optimize the functional utility of the learned language. In Fig.~\ref{fig:captions}, we see that optimizing this objective can indeed improve the capability of the language encoder. The location of the object in the image cannot be clearly distinguished via supervised learning using human samples, but can be captured after optimizing the mutual information term. Intuitively, when the language encoder cannot describe locations, the action distribution of the listener would not change after reading in the words describing the location, resulting in a lower mutual information value. Therefore, maximizing MI can force the encoder to distinguish the location information.

As described in Sec.~\ref{sec:le}, apart from the mutual information term, we also train the language encoder by rewards given by the environment. We present learning curves in Fig.~\ref{fig:comp-1} (c) as ablation studies to show the contribution of these two terms. We can observe that simultaneous optimization of mutual information and rewards has the best performance, while solely optimizing the mutual information is the worst, even worse than the language encoder trained only by human samples. This is because mutual information is large as long as the messages can influence the action selection of the listener, even when such influence is detrimental to task performance. Augmenting the learning objective with rewards can solve this problem -- the mutual information now tends to positively influence the listener's policy.

We now demonstrate that the mutual information can serve as a standard for learning a economy of language on a given task. In Fig.~\ref{fig:word_mi}, we show the mutual information between each word and the listener's action selection (Eq.~\ref{equ:functional_obj}). In the first example, in order to distinguish the target image from the distracting image, the speaker only needs to convey information about the object's color. Our language encoder reaches our expectation, with the word 'purple' having the largest mutual information. The location is also a distinguishing feature, and the words 'bottom' and 'right' also have a high value of mutual information. Similar conclusions can be drawn for the rest examples in Fig.~\ref{fig:word_mi}.

\begin{figure}
    \centering
    \includegraphics[width=\linewidth]{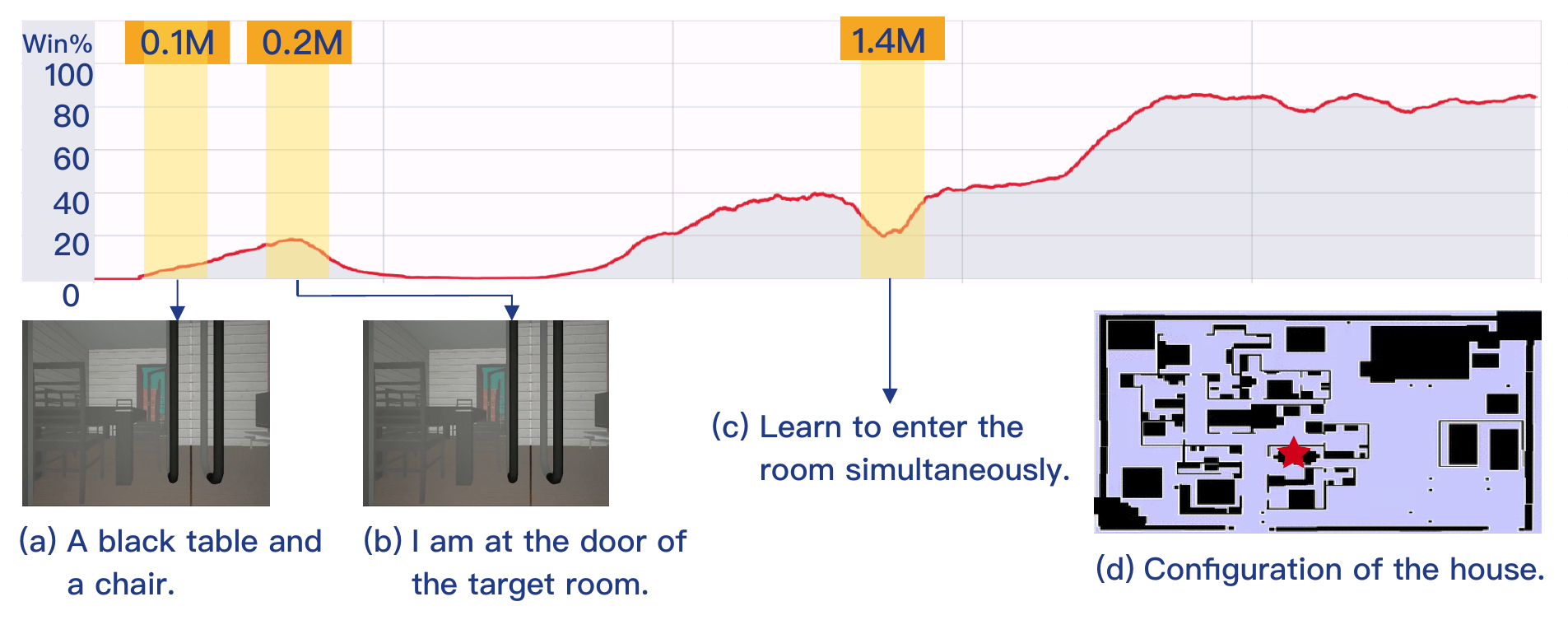}
    \caption{Learning curves and three critical phases during the training process of our method on the 3D navigation game. (a) The initial language model gives a description which cannot distinguish different rooms. (b) After trained with human samples, the mutual information and reinforcement learning objective, agents do not describe the details of the room anymore. (c) Agents gradually learn to coordinate their actions of entering the room based on the receipt of the messages. }
    \label{fig:house}
\end{figure}
\subsection{3D Navigation Game}
In Fig.~\ref{fig:house}, we show results on the 3D navigation game. In a one-floor house shown in Fig.~\ref{fig:house} (d), two agents need to enter the target room, marked by the star, simultaneously. The initial pre-trained language model outputs "a black table and a chair" when observing the target room. Although this description makes sense and is correct, many rooms in the house have a black table and chair. Consequently, this model cannot distinguish the target room from the other rooms. After 200k interactions with the environment, our model learns to ignore the details when describing the room. Rather, it summarizes the visual inputs and directly tells other agents whether the ego agent is at the door of the target room. Although the communication protocol is near-optimal at this point, agents haven't learned to coordinate their actions according to the received messages. Policy learning and language decoder learning of action coordination takes about 2M samples. Agents gradually learn to enter the room when they say and hear the message, "I am at the door of the target room."
\section{Closing Remarks}
In this paper, we study how to improve the performance of spoken dialogue interfaces. We decompose the uncertainty of spoken language into an entropy term measuring the structural uncertainty and a mutual information term reflecting the functional utility of the message. By optimizing these objectives within a reinforcement learning framework, for unseen tasks, our method can quickly learn task-specific communication protocols in natural language, which can help agents and human experimenters finish complex, temporally extended tasks.

A drawback of our method is that the learned language is not perfectly precise. For example, it can not distinguish the shape of the object in the referential game. Moreover, on the 3D navigation task, communication among agents is noisy, occasionally sending messages describing connectivity of the house map or details of visual observation. We plan to look into these issues in the future.


\bibliographystyle{ACM-Reference-Format}
\bibliography{sample-base}

\appendix

\end{document}